\documentclass{aastex}
\usepackage{spr-astr-addons}
\usepackage{url}

\begin{document}

\title{No uniform density star in general relativity} 
\shorttitle{No uniform Density Star}
\shortauthors{A. Mitra}

\def\b{\begin{equation}}
\def\e{\end{equation}}
\def\l{\left}
\def\r{\right}

\author{Abhas Mitra\altaffilmark{1}}

\altaffiltext{1}{Theoretical Astrophysics Section, Bhabha Atomic Research Centre,
    Mumbai -400085, India: Email: amitra@barc.gov.in}

\begin{abstract}
As per   general relativity (GR), there cannot be any superluminal propagation of energy.
And thus, the sound speed in a continuous medium, $c_s=\sqrt{dp/d\rho}$,  must be subluminal. However, if one would conceive of a {\em homogeneous} fluid, one would have $c_s=\infty$ unless pressure too would be homogeneous. Thus it is universally accepted that 
the maiden   GR interior solution obtained by Schwarzschild,
 involving a homogeneous fluid having a boundary, is unphysical.  However no one has ever shown how this exact solution  is in reality  devoid of physical reality. Also,
this solution is  universally used for approximate modelling of general relativistic stars and compact objects. 
But here first we show that in order that the Kretschmann scalar is continuous, one should have $\rho=0$ for strictly homogeneous static stars. Further, by invoking the fact that in GR, given one time label $t$ one can choose another time label $t_*=f(t)$ {\em without any loss of generality}, we obtain the same result that for a static homogeneous sphere $\rho=0$.  Consequently, it is eventually found that the static homogeneous sphere having a boundary is just part of the vacuum where $c_s=0$ rather than $\infty$. Therefore all general relativistic stars must be inhomogeneous.
\end{abstract}


-----------






 
\section{Introduction}
It is well known that  general relativity prohibits superluminal motion which
can be associated with flow of matter or energy momentum. And therefore it definitely prohibits occurrence of superluminal sound speed $c_s = \sqrt{dp/d\rho}$, where $p$ is the isotropic pressure and $\rho$ is the density of the fluid. But if one would conceive of a homogeneous fluid having $d\rho=0$, one would immediately have $c_s=\infty$. Therefore, unlike Newtonian physics, GR prohibits, density homogeneity unless there is a homogeneity of pressure too so that both  $dp=d\rho =0$.

But, immediately after the formation of general relativity (GR) Schwarzschild came out with the maiden {\em interior} solution\citep{b6}. He considered a simple case of a static spherically symmetric fluid sphere of uniform density $\rho$. Since then almost every general relativist has studied and reconsidered this important solution. The most important thing about this solution is that it is singularity free and has  served as a model of general relativistic stars  ever since, i.e., for 94 years.
Given a spherically symmetric metric ($G=c=1$):
\b
ds^2 = e^\nu dt^2 - e^\lambda dr^2 - r^2 d\Omega^2
\e 
where $d\Omega^2= d\phi^2 + \sin^2 \theta d\theta^2$, 
the interior Schwarzschild solutions is well known\citep{b3,b6,b7,b8}:
\b
e^{\nu/2} = {1\over 2} \l[ 3 \sqrt{1 - 8\pi\rho R^2 /3} - \sqrt{1- 8 \pi \rho r^2 /3}\r]
\e
and
\b
e^{-\lambda/2} = \sqrt {1-8\pi \rho r^2/3}
\e
where the boundary of the fluid $r=R$ is defined from the condition that pressure vanishes there $p=p_b =0$. 

But since in this case $d\rho=0$ and {\em apparently}, $dp \neq 0$, the Schwarzschild solution certainly defies the GR diktat that $c_s <1$. Therefore this solution, must eventually, be devoid of physical reality, i.e; the ``homogeneous fluid having a {\em finite} density and a boundary'' must be illusory. While every author acknowledges this, nobody has shown why the solutions (2) and (3) must actually  be an illusion devoid of physical reality.

To resolve this paradox, we show below that the only constant density fluid the Schwarzdchild sphere can physically signify is vacuum having $p=\rho=0$, where one has $c_s =0$ rather than $c_s=\infty$.

For a proper appreciation of this {\em subtle} proof, we need to first visit, the first vacuum solution of GR, namely, the so-called vacuum Schwarzschild solution.

\subsection{Spherically Symmetric Vacuum Solution}
As one solves the Einstein equations for spherically symmetric {\em vacuum} spacetime, one {\em initially} obtains the following metric \citep{b4}
 \b
ds^2 = e^{h(t)} (1- \alpha_b/r) dt^2 - (1 - \alpha_b/r)^{-1} dr^2 - r^2 d\Omega^2
\e 
where, to begin with,  both $h(t)$ and $\alpha_b$ are unknown integration constants. However, since these integration constants are obtained from the solution of Einstein equations, in general, they may involve parameters like $p_0, \rho_0,  M_b$ etc. where the subscript $0$ denotes central values, $b$ denotes boundary values and $M$ denotes gravitational mass. And naturally, they cannot involve parameters not {\em appearing in the relevant Einstein equations at all such as proton or electron mass, Planck's constant, Fine structure constant} etc. In particular, in this given case, since $h=h(t)$, it can  depend neither on any fundamental constant nor on any time independent parameters like $p_0, \rho_0, M_b$ etc.

Given this obvious limitation,  in GR, given one time label $t$,
one can always select another time label\citep{b4}
\b
t \to t_* = f(t)
\e
without affecting {\em the physical content} of the problem. Thus one can choose a new label such that
\b
dt_*^2 = e^h dt^2
\e
In this new time label, the metric becomes
\b
ds^2 =  (1- \alpha_b /r) dt_*^2 - (1 - \alpha_b /r)^{-1} dr^2 - r^2 d\Omega^2
\e 
It is customary here to drop the asterisk, and rewrite the above metric as
\b
ds^2 =  (1- \alpha_b /r) dt^2 - (1 - \alpha_b /r)^{-1} dr^2 - r^2 d\Omega^2
\e
This effectively means that in the new time label, one has $e^h =1$. Now, if one would demand that the spacetime must be Newtonian at very large $r$, one would be able to identify $\alpha_b= 2M_b$:
\b
ds^2 =  (1- 2M_b/r) dt^2 - (1 - 2M_b/r)^{-1} dr^2 - r^2 d\Omega^2
\e
where $M_b$ is the gravitational mass of the spherical region beyond which the vacuum solution is valid.

Thus we found that the coordinate freedom of choosing an arbitrary time label is necessary to attribute physical meaning to the vacuum Schwarzschild solution; i.e., for identifying the constant $\alpha_b$ as proportional to the gravitational mass.

Now we shall find below that a similar application of this GR principle of coordinate freedom is necessary to instil physical reasonableness for the interior (homogeneous) Schwarzschild solution too. 
\section{Matching of Interior and Exterior Solutions}
It is well known that the interior solutions (2) and (3) match the exterior solution (9) because
\b
M_b = {4\pi \rho\over 3} R^3
\e
However, proper matching of any interior solution with an exterior one demands not only matching of $g_{ab}$ at the boundary but matching of the various derivatives of $g_{ab}$ too. And the minimum requirement is that atleast the first order derivatives should match at the interface\citep{b2, b5}.
From equations (2) and (3) that, for the interior metric, we have
\b
g_{rr}' = {16 \pi  \rho r \over  3(1- 8\pi r^2\rho/3)^2}
\e
\begin{eqnarray}
g_{00}' = && {4\pi \rho r \over 3 \sqrt{1- 8\pi \rho r^2/3}} \times \nonumber \\
&& \l[ 3 \sqrt{1 - 8\pi\rho R^2 /3} - \sqrt{1- 8 \pi \rho r^2 /3}\r]
\end{eqnarray}
where the prime denotes differentiation by $r$. On the other hand, for the exterior metric, one has
\b
g_{rr}' = {2M_b\over r^2 (1- 2M_b/r)^2 }
\e
and
\b
g_{00}' = {2M_b\over r^2}
\e
 And the interior constant density solutions (2) and (3) violate this condition because while from Eq.(11) one obtains
\b
g_{rr}'|_i = {(16 \pi \rho R/3)\over (1- 8\pi \rho R^2/3)^2}
\e
from Eq.(13), one finds
\b
g_{rr}'|_e = {(8 \pi \rho R/3)\over (1- 8 \pi \rho R^2/3)^2}
\e
i.e.,
from Eq.(11) one obtains
\b
g_{rr}'|_i = 2 g_{rr}'|_e
\e
Therefore the interior and exterior solutions really do not gel together  {\em if one would assume} $\rho >0$. Consequently, mere existence of solutions (2) and (3) need not necessarily imply that $\rho=constant >0$ for this problem. 

Nonetheless, Following Robson, one can attempt matching of the first derivatives of $g_{rr}$, by changing the radial variable in the exterior section, to another appropriate variable\citep{b5} 
\b
r_* = r + A (r-R)^2
\e
where the constant
\b
A= {3M_b\over 2R^2 (1- 2M_b/R)}
\e
In this new variable, one obtains
\b
g_{r_* r_*} = -(1- 2M_b/r)^{-1} [1+ 2A (r-R)]^{-2}
\e
and
\b
{d g_{r_* r_*}\over d r_*} =  {4 M_b\over R^2 (1-2M_b/R)^2}; \qquad r=R
\e
Now one can ensure the matching of the respective first derivatives at $r=R$:
 \b
{d g_{r_* r_*}\over d r_*}|e =  {d g_{rr}\over dr}|i = {(16 \pi \rho R/3)\over (1- 8\pi \rho R^2/3)^2}; \qquad r=R
\e

  What is however overlooked here is that such a coordinate transformation is unphysical because it follows that
\b
-g_{r_* r_*} \to 0; ~as \qquad r, r_*\to \infty; ~if~ indeed ~A >0
\e
i.e., the metric would display a most unphysical singularity at $r=r_*=\infty$ if indeed one would have $\rho >0$! Thus, matching of the interior solution with such an unphysical exterior solution would mean nothing physically meaningful.

On the other hand,  one can ensure such matching {\em in an invariant} way and without resorting to any unphysical coordinate transformations only by setting $\rho=constant =0$.
Let us see below that why indeed one should have $\rho =0$ in this problem.

\section{Continuity of Spacetime Curvature}
Physically valid solutions must ensure that there is no discontinuity in the spacetime curvature at the interface of interior and exterior solutions. 

And given a static metric of the form (1), the Kretschmann scalar is given by\citep{b1}
\b
{\cal K}= K_1^2 + 2K_2^2 +2K_3^2 + 4K_4^2
\e
where
\b
K_1 = e^{-\lambda} (2\nu'' +\nu'^2 - \lambda' \nu'),
\e
\b
K_2 = e^{-\lambda} {\nu'\over r},
\e
\b
K_3 = {-\lambda'\over r} e^{-\lambda},
\e
\b
K_4 = {1- e^{-\lambda}\over r^2}
\e
In order to calculate the Kretschmann scalar easily, we find a simplified form of the interior metric. First, let us recall the local energy conservation equation\citep{b4}
\b
\nu' = - {2p'\over p+\rho}
\e
 For a {\em uniform} $\rho$, this can be easily integrated to obtain
\b
\nu(r) - \nu_0 = \ln \l({p_0 +\rho \over p + \rho}\r)^2
\e
 where $0$ denotes central values. From the foregoing equation, one finds
\b
e^\nu = \l({p_0 +\rho\over p + \rho}\r)^2~e^{\nu_0}
\e
Therefore, one finds that the interior metric has a form
\b
ds^2 =  \l({p_0 +\rho\over p + \rho}\r)^2 e^{\nu_0} dt^2 - e^\lambda dr^2 - r^2 d\Omega^2
\e
where $e^\lambda$ is still given by Eq.(3).

Now we recall the Tolman -Oppenheimer - Volkoff equation:

\b
p' = - {(\rho +p) (M + 4 \pi r^3 p)\over  r^2 (1 - 2M/r)}
\e
where, for the uniform density case
\b
M = {4\pi\over 3} \rho r^3
\e
and
\b
p' = - {4\pi r\over 3} {(\rho +p) (\rho + 3 p)\over  (1 - 8\pi \rho r^2/3)}
\e
By using the foregoing equation in (28), we see that
\b
\nu' = {8\pi r\over 3} { (\rho + 3 p)\over  (1 - 8\pi \rho r^2/3)}
\e

Now, after a rather lengthy calculation, one can find the expression for the Kretschmann scalar for the interior solution:
\b
{\cal K}^i = {256\pi^2\over 3} [\rho^2 + (\rho+3p)^2]
\e
On the other hand, for the exterior vacuum metric, one finds that
\b
{\cal K}^e = 48 {M_b^2\over r^3}
\e
By using equations (10),  (36) and (37), it is found that, at the boundary $r=R$ where $p=0$, {\em there will be a discontinuity in the spacetime curvature} if indeed $\rho >0$.
\b
{\cal K}^i = 2 {\cal K}^e; \qquad r=R; ~if ~ \rho>0
\e
Therefore, ${\cal K}^i$ and ${\cal K}^e$ {\bf differ by a factor} of $2$ at the boundary. Earlier, we noted a discrepency by the same factor of $2$ in Eq.(17).
However spacetime curvature must be unique at a given spacetime point; i.e., we must have
${\cal K}^i={\cal K}^e$ at $r=R$. And this can be ensured by realizing that {\em the only constant density static sphere allowed by GR is the vacuum}; i.e., $\rho=0$. When $\rho=0$, one should naturally get $p=0$ too.

\section{Metric at the Centre of Symmetry}
In general, by the principle of equivalence, at a given spacetime point, one can bring any metric at a Minkowskian form. Obviously, that does not mean that the spacetime is flat at that point.  Having remembered this basic lesson,
let us note that at $r=0$, even for an inhomogeneous case, the metric (1) would assume a form
\b
ds^2 \to e^{\nu_0} dt^2 - e^{\lambda_0} dr^2 - r^2 d\Omega^2
\e

Since by definition $\nu_0 \ne \nu_0(r)$, by using the coordinate freedom in choosing an arbitrary time label, and following Eqs.(5) and (6),
 let us choose a new time label $t_*$ such that
\b
dt_* = e^{\nu_0/2} ~ dt
\e
In this new time label, at $r=0$, we have
\b
ds^2 \to dt_*^2 - e^{\lambda_0} dr^2 - r^2 d\Omega^2
\e
For an inomogeneous case, $e^{\lambda_0} \neq 1$, and the metric continues to be non-flat at $r=0$. Even if we would make a transformation
\b
r_* = r e^{\lambda_0/2}
\e
we would obtain
\b
ds^2 \to dt_*^2 - dr_*^2 -  e^{-\lambda_0} r_*^2 d\Omega^2
\e
and which is not the correct Minkowski form (in spherical polar coordinates).

However, for the homogeneous case, we find from Eq.(3) that
\b
e^{\lambda_0} =1
\e
and, at $r=0$, the metric appears to become naturally flat
\b
ds^2 \to  dt_*^2 - dr^2 - r^2 d\Omega^2
\e
Therefore we expect the Kretschmann scalar at $r=0$ 

\b
{\cal K}^i={\cal K}_0 = {256\pi^2\over 3} [\rho^2 + (\rho +3p_0)^2]
\e
to vanish.
And one can have $K_0 =0$, only when $\rho=p_0=0$. Thus we again see that the uniform density static sphere to be just the vacuum!

\section{Confirmation by Different Route}
We find that ${\cal K}^i$ does not depend on the value of $e^{\nu_0}$ at all! Thus {\em irrespective of any other details}, one can set it to any value one wishes; and what is important here is that exercise of this freedom does not affect the intrinsic value of ${\cal K}$ or any other relevant physical aspects. This is actually guaranteed by the coordinate freedom (5) and (6). Now note from Eq.(2) that

\b
e^{\nu_0/2} = {1\over 2} \l( 3\sqrt{1- 8\pi \rho R^2/3} -1\r)
\e

and
\b
e^{\nu_b/2} =  \sqrt{1- 8\pi \rho R^2/3} 
\e

And by combining these two equations we find that 
\b
e^{\nu_0/2} = {1\over 2} \l( 3e^{\nu_b/2} -1\r)
\e

On the other hand, from Eq.(30), we see that
\b
e^{\nu/2} = \l({p_0 +\rho\over p + \rho}\r)~e^{\nu_0/2}
\e
By setting $p=0$, the above equation, we find that
\b
e^{\nu_b/2} = \l({p_0 +\rho\over  \rho}\r)~e^{\nu_0/2}
\e
Then using Eq.(51) on the right hand side of the foregoing equation, we have
\b
e^{\nu_b/2} = \l({p_0 +\rho\over  2\rho}\r)~\l[ 3e^{\nu_b/2}-1\r]
\e
Now  equations (49) and (52) can be solved to obtain

\b
e^{\nu_b/2} = \l({p_0 +\rho\over  \rho +3p_0}\r)
\e
and
\b
e^{\nu_0/2} = { \rho\over \rho + 3p_0}
\e

As we have already noted, by invoking the coordinate freedom in choosing the time label in GR, one can set arbitrary fixed value of $\nu_0\neq \nu_0(r)$ without any loss of generality. Thus in the new time label $t_*$  we have

\b
ds^2 =  \l({p_0 +\rho\over p + \rho}\r)^2  dt_*^2 - e^\lambda dr^2 - r^2 d\Omega^2
\e
This means that, in the new time label one has
\b
e^{\nu_0/2} =1
\e
i.e., in this new time label, one has
\b
{\rho \over \rho + 3p_0} =1
\e
just like in a new time label we had $e^h =1$.
Now from the above equation,  it is seen that, {\em in this new time label}, one has
\b
p_0 =0
\e

But the central pressure can vanish only when the density vanishes. Therefore, in this new time label one has both $p=\rho=0$. However since both $p$ and $\rho$ are scalars, they must be independent of time labels. In other words, the freedom of exercise of the coordinate freedom demands that, for this problem, one must intrinsically have $\rho =p=0$ in accordance with the result obtained in the previous section.

It is also important to note that {\em had the density been position dependent, we would not have been able to set it to zero} by invoking the freedom (5).

\section{Discussions}
The maiden and important interior solution of Schwarz- schild has been used by all relativists as a basic model for stars in general relatvity. This has happened even when most of the athors have acknowledged the fact that a {\em strict} uniform density must not be realizable because this would imply the occurrence of not only a superluminal but an infinte sound speed. Yet this solution has been considered as a good approximation for the physical reality which apparently shows that $2M_b/R <8/9$, where $M_b$ is the gravitational mass of the star having a radius $R$.  While we cannot go here to thousands textbooks and articles dealing with this problem, we may quote from the first comprehensive GR text book by \citep{b3}:

\begin{verbatim}
 * Schwarzschild's solution is of 
considerable interest;
\end{verbatim}

Such an idea, that the Schwarzschild solution is {\em interesting}, though idealistic, has
been adopted by every author. For instance, consider the comments by \cite{b8} (see pp.330) in this context:

\begin{verbatim}
 General relativity finds an interesting 
application to one other class of stable
 stars, those consisting of incompressible
 fluids, with an equation of state
\end{verbatim}
\b
\rho =constant
\e
 But here we  resolved the paradox of how GR can {\em apparently} allow meaningful solutions (2) and (3) which correspond to infinite sound speed.  To be more precise, we resolved this paradox by showing

$\bullet$ In order that the spacetime is continuous at the boundary of the supposed constant density star, one must have $\rho=0$.

Later, 

$\bullet$ we obtained the same result independently by invoking
  the fundamental GR principle that different time labels selected as per Eq.(5) must correspond to the same physical reality. 

 Thus we found that, GR enforced $c_s=0$ for a situation which has  so far been considered to lead to $c_s=\infty$ in blatant violation of GR. Finally, the paradox was occurring when one was missing out on something {\em very subtle yet very fundamental}. And here the subtle and fundamental point was that
physical results must be invariant under the transformation (5). In fact all paradoxes of physics have a similar origin: they appear only because one misses out on something very important which is {\em so subtle} that it gets overlooked!

Therefore, {\em for the first time}, we have proved here that all general relativistic stars must be non-uniform.


\begin{thebibliography}{99}
\bibitem[\protect\citeauthoryear{Bronnikov \& Kim}{2003}]{b1} Bronikov, K.A. and Kim, S-W. 2 Phys. Rev. D. 67, 064027, (2003)

\bibitem[\protect\citeauthoryear{Dandach}{1992}]{b2} Dandach, N.F., IL Nuovo Cimento, 107B(11), 1267, (1992)
\bibitem[\protect\citeauthoryear{Eddington}{1923}]{b3} Eddington, A.S., {\it The Mathematical Theory of Relativity}  (Cambridge Univ. Press, Cambridge, 1923)

\bibitem[\protect\citeauthoryear{Landau \& Lifshitz}{1962}]{b4} Landau, L.D. and Lifshitz, E.M., {\it Classical Theory of Fields}, (Pergamon Press, Oxford, 1962)

\bibitem[\protect\citeauthoryear{Robson}{1972}]{b5} Robson, E.H., Ann. Henri Poincare, 16, 41, (1972)
\bibitem[\protect\citeauthoryear{Schwarzschild}{1916}]{b6} Schwarzschild, K., Sitz. Preuss. Acad. Wiss., 424, (1916)


 \bibitem[\protect\citeauthoryear{Tolman}{1934}]{b7} Tolman, R.C., {\it Relativity, Thermodynamics \&
Cosmology}, (Clarendon Press, Oxford, 1962)

\bibitem[\protect\citeauthoryear{Weinberg}{1972}]{b8} Weinberg, S., {\it Gravitation and Cosmology: Principles and Applications of General Theory of Relativity}, (John Wiley, New
York, 1972)



\end{thebibliography}
\end{document}